\documentclass[a4paper,fleqn,usenatbib]{mnras}

\usepackage[T1]{fontenc}
\usepackage{ae,aecompl}

\usepackage{graphicx}	
\usepackage{amsmath}	
\usepackage{amssymb}	

\def \rev  { }

\title[Abell 1201 counter-image]
{A 
counter-image to the gravitational arc in Abell 1201:
Evidence for 
IMF variations,
or a 10$^{10}$\,M$_\odot$ black hole?\thanks{Based on 
observations collected at the European Southern Observatory, Chile (ESO Programme 077.A-0806(A)).}\thanks{Based on observations made with the NASA/ESA Hubble Space Telescope, obtained from the Data Archive at the Space Telescope Science Institute, which is operated by the Association of Universities for Research in Astronomy, Inc., under NASA contract NAS 5-26555. These observations are associated with program 08719.}}
\author[Russell J. Smith et al.]{
Russell J. Smith\thanks{E-mail: russell.smith@durham.ac.uk}, John R. Lucey and Alastair C. Edge
\\
Centre for Extragalactic Astronomy, University of Durham, Durham DH1 3LE\\
}

\pubyear{2016}

\begin{document}
\label{firstpage}
\pagerange{\pageref{firstpage}--\pageref{lastpage}}
\maketitle

\begin{abstract}
Abell 1201 is a massive galaxy cluster at z=0.169 with a brightest cluster galaxy (BCG) that acts as a gravitational lens to a background source at z=0.451. The lensing configuration is unusual, with a single bright arc formed at small radius ($\sim$2\,arcsec), where stars and dark matter are both expected to contribute substantially to the total lensing mass. Here, we present deep spectroscopic observations of the Abell 1201 BCG with MUSE, which reveal emission lines from a faint counter-image, opposite to the main arc, at a radius of 0.6\,arcsec. We explore models in which the lensing mass is described by a combination of stellar mass and a standard dark-matter halo. The counter-image is not predicted in such models, unless the dark-matter component is negligible, which would imply an extremely heavy stellar initial mass function (IMF) in this galaxy. We consider two modifications to the model which can produce the observed configuration without resorting to extreme IMFs. Imposing a radial gradient in the stellar mass-to-light ratio, $\Upsilon$, can generate a counter-image close to the observed position if $\Upsilon$ increases by $\ga$60 per cent within the inner $\sim$1\,arcsec (e.g. variation from a Milky-Way-like to a Salpeter-like IMF). Alternatively, the counter-image can be produced by introducing a central super-massive black hole. The required mass is
$M_{\rm BH}$\,=\,(1.3$\pm$0.6)$\times$10$^{10}$\,M$_\odot$, which is comparable to the largest black holes known to date, several of which are also hosted by BCGs. 
We comment on future observations which promise to distinguish between these alternatives.
\end{abstract}


\section{Introduction}

Rich galaxy clusters are extreme locations: they occupy the largest dark-matter haloes in the universe, and harbour the 
most massive galaxies at their centres. In turn, the central galaxies of massive clusters host some of the most extreme black holes 
known to date.

Clusters are dominated by dark matter (DM) at all but the smallest radii, so they provide important constraints on the structures of 
DM haloes. At large radii, weak-lensing and X-ray data largely support the \citet*{1995MNRAS.275..720N} (NFW) functional form 
\citetext{e.g. \citealt{2003ApJ...598..804K}; \citealt{2007MNRAS.379..209S}}. 
The slope of the DM profile towards the cluster centre is 
sensitive to the micro-physics of the DM particle \citep{2000PhRvL..84.3760S},  as well as to interactions
between baryonic and dark components \citetext{e.g. \citealt{1986ApJ...301...27B}}. 
Determining the inner halo profile slope is, however, hampered by the presence of the brightest cluster galaxy (BCG), located at or near
the halo centre. Within a radius of a few kpc, the stellar mass density of the BCG is comparable to, or exceeds, the DM density. 
Hence the observational challenge of studying the central structure of the DM halo is coupled to that of understanding the stellar component
\citep{2004ApJ...604...88S}.

The BCG stellar mass contribution is also a matter of interest in its own right, since
massive elliptical galaxies are widely suspected to harbour stars formed according to an
initial mass function (IMF) different from that pertaining to the Milky Way (MW)
\citetext{e.g. \citealt{2010ApJ...709.1195T,2012ApJ...760...71C,2012Natur.484..485C}}.
If the IMF variations are associated with the physical conditions in violent starburst events at early epochs  \citetext{e.g. \citealt*{2014ApJ...796...75C}}, 
then the centres of BCGs are a likely site to habour the affected populations.
In a recent study using a combination of stellar dynamics and gravitational lensing constraints on the mass profile of BCGs, 
\citet{2013ApJ...765...25N} found a preference for both shallower-than-NFW DM profiles and heavier-than-MW IMFs, on average.

Finally, the most massive galaxies, in the most massive haloes, are also likely hosts for the largest central black holes (BH) in the universe
\citep{2011Natur.480..215M,2012MNRAS.424..224H}.
In distant lensing clusters, kinematic data do not resolve the dynamical effects of BHs, and the relative contribution of the BH to the lensing 
mass is usually negligible. However, for clusters at lower redshift, sufficiently massive BHs may have measurable effects on the
stellar kinematics at small radius. For certain configurations, massive BHs can also affect the lensing caustic structure, altering the number of 
images observable \citep*{2001MNRAS.323..301M}.

In this paper, we present the first results from new wide-field integral-field spectroscopic observations of the $z$\,=\,0.169 cluster Abell 1201. 
\cite{2003ApJ...599L..69E} (hereafter E03) identified a bright tangential arc around the BCG using shallow
{\it Hubble Space Telescope} imaging with WFPC2 (Wide Field and Planetary Camera 2), 
obtained as part of a systematic search for lensing clusters \citep{2004ApJ...604...88S}. 
The lensing configuration of Abell 1201 is unusual, in that the arc is located at a radius of only $\sim$2\,arcsec ($\sim$6\,kpc), well 
within the effective radius of the BCG, rather than at the $\sim$10\,arcsec scales typical for cluster lenses. 
E03 also presented Keck spectroscopy from which they measured a redshift of $z$\,=\,0.451 for the arc.
X-ray observations of Abell 1201 \citep{2009ApJ...692..702O,2012ApJ...752..139M}  indicate a 
post-merger morphology for Abell 1201, with the merger direction aligned with the BCG major axis, 
and the BCG itself offset from the X-ray peak by $\sim$11\,kpc along the same axis. From the radial velocities of 165 member galaxies, 
\citet{2013ApJ...767...15R} measure a cluster velocity dispersion $\sigma_{\rm cl}$\,=\,$683^{+68}_{-53}$\,km\,s$^{-1}$, and derive a virial
{\rev mass $M_{200}$\,=\,(3.9$\pm$0.1)$\times$$10^{14}$\,M$_{\odot}$ (for $h$\,=\,0.678)} from the infall caustic fitting method.
Based on Sloan Digital Sky Survey (SDSS) photometry, the BCG has a luminosity of $L_r$\,$\approx$\,4$\times$10$^{11}$\,L$_{\odot,r}$,   
while the Two Micron All Sky Survey  \citep{2000AJ....119.2498J} yields $L_K$\,$\approx$\,1.6$\times$10$^{12}$\,L$_{\odot,K}$. The BCG
has a half-light radius of $r_{\rm eff}$\,$\approx$\,15\,kpc. 
\citet{2004ApJ...604...88S} measured a velocity dispersion of $\sigma$\,=\,230--250\,km\,s$^{-1}$ in the inner 1.5\,arcsec; SDSS reports 
$\sigma$\,=\,277$\pm$14\,km\,s$^{-1}$. The luminosity, radius and velocity dispersion for the Abell 1201 BCG are consistent with the 
early-type galaxy Fundamental Plane. 

Our new integral-field  observations were motivated by the unusually small separation of the bright arc in Abell 1201, which makes it feasible to 
combine stellar kinematics and strong-lensing information across an overlapping range in radius, which is not possible in most lensing clusters. 
The velocity dispersion measurements from the \citet{2004ApJ...604...88S} long-slit spectra do not reach the radius of the arc in Abell 1201, while
previous integral-field observations \citep{SwinbankThesis} covered a much narrower field-of-view 
and sampled a limited spectral range, not including the bright [O\,{\sc iii}] emission line from the arc.
Our stellar kinematic measurements and dynamical modelling will be presented in a forthcoming paper. 
Here, we focus on the  strong-lensing constraints, showing that
the unusual configuration of Abell 1201 allows us to infer the presence of an additional centrally-concentrated mass, 
of order $10^{10}$\,M$_\odot$, with no detectable luminous counterpart. 

The remainder of the paper is structured as follows: Section~\ref{sec:data} describes the observations and data reduction steps, and presents the 
general lensing configuration, including identification of a faint counter-image close to the centre of the BCG. 
Section~\ref{sec:mainmod} presents a lensing analysis constrained only by the main arc, using models with a constant stellar mass-to-light ratio
combined with a parametrized dark-matter halo. 
Section~\ref{sec:count} then discusses the interpretation of the counter-image, proposing three alternative scenarios: 
(a) a very heavy IMF throughout the BCG, (b) a steep radial variation in the stellar initial mass function, and (c) a very massive central black hole. 
In Section~\ref{sec:disc} we discuss the merits and implications of these solutions, with reference to external evidence, and Section~\ref{sec:outlook}
considers future observations which might help discriminate between them. Brief conclusions are summarized in Section~\ref{sec:concs}.

For computing physical scales we adopt the relevant cosmological parameters from  \cite{2016A&A...594A..13P}:
$h$\,=\,0.678, $\Omega_{\rm M}$\,=\,0.308 and $\Omega_{\rm \Lambda}$\,=\,0.692. In this cosmology, the spatial scale at the 
redshift of Abell 1201 is 2.96\,kpc\,arcsec$^{-1}$.

\section{MUSE observations}\label{sec:data}

\begin{figure*}
\includegraphics[width=180mm]{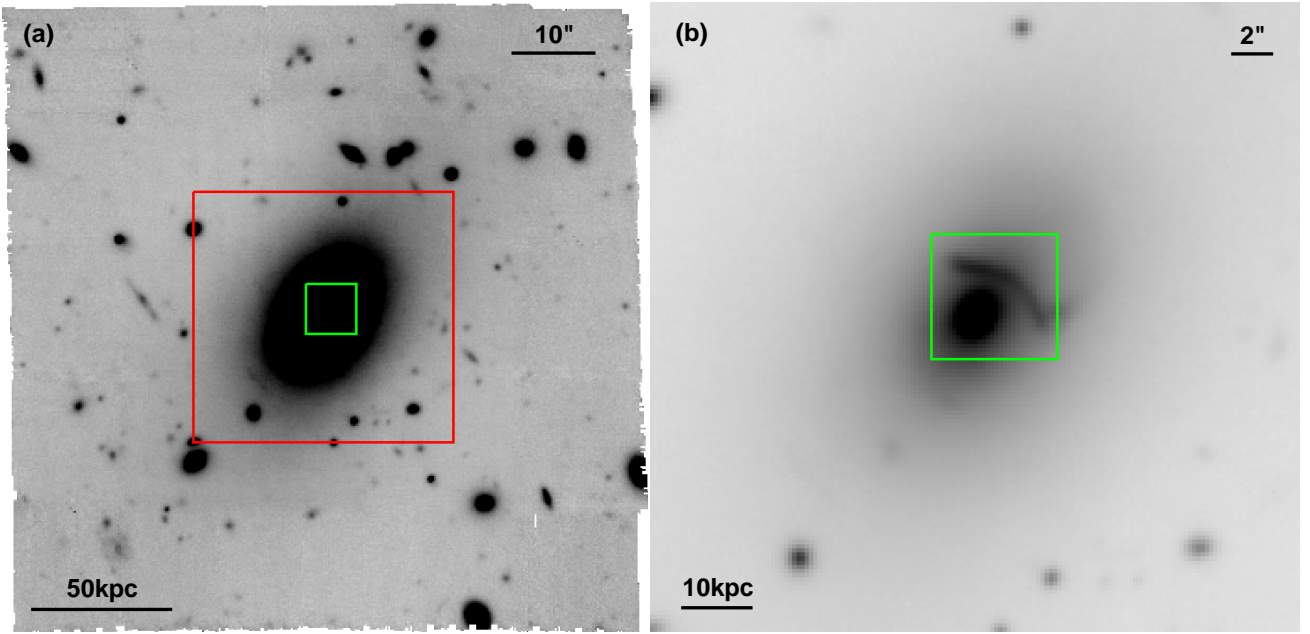}
\vskip -2mm
\caption{Collapsed image of the Abell 1201 field, from the final combined MUSE
datacube, over the wavelength interval 6600--7600\,\AA. Panel (a) shows the full field-of-view with the grey-scale optimised to show faint galaxies
and the outer parts of the BCG. The red square indicates the region expanded in Panel (b), in which the grey-scale is scaled to show the 
continuum light from the main arc. In both panels, the green square indicates the 6$\times$6\,arcsec$^2$ area depicted in 
Figures~\ref{fig:arcvel_map},~\ref{fig:lensmods_main},~\ref{fig:varml},~\ref{fig:lensmods_coun} and \ref{fig:velpred}.
}
\label{fig:img}
\end{figure*}

We observed Abell 1201 with the Multi-Unit Spectroscopic Explorer (MUSE) \citep{2010SPIE.7735E..08B} on the 
8.2m Yepun (Unit Telescope 4) of European Southern Observatory's Very Large Telescope. 
The data used in this paper were obtained on the nights of
March 31 and April 2, 2016, under good seeing conditions. 

A total of twelve 940-second exposures were obtained, using the standard spectral configuration, covering 4750--9350\,\AA, sampled at 1.25\,\AA\ per pixel,
with resolution 2.6\,\AA\ FWHM (at $\lambda$\,=\,7000\,\AA). Each exposure spans a $\sim$1\,arcmin$^2$ field-of-view, with 0.2\,arcsec spatial pixels. 
To  help suppress the effects of instrumental artifacts, the observations were arranged in four groups of three exposures each. Each group was observed 
at a different position angle (0, 90, 180, 270 degrees),
and the field centres for the groups were arranged in a 2$\times$2 grid, with separation of 15\,arcsec. Hence the total field observed is 75$\times$75\,arcsec$^2$, 
while the full exposure time of 3.1\,hours was obtained only in the central 45$\times$45\,arcsec$^2$. 
Further small dither offsets ($\sim$0.5\,arcsec) were made between the exposures in each group. 

The initial data reduction steps were accomplished using the standard MUSE pipeline. Each of the twelve exposures was 
reconstructed to generate a separate datacube, using an initial ``global'' sky spectrum obtained from the darkest parts from the complete field of view.
This leaves a wavelength-dependent background ``striping'' effect, apparently due to residual bias-level variations which differ among
the 24 separate spectrograph ``channels'' of the instrument. To reduce the impact of these variations, we derived and subtracted a 
separate ``residual sky'' spectrum from the darkest spatial pixels in each {\it channel}, prior to combining the separate observations into a 
single final data-cube. During this final combination, integer-pixel astrometric offsets were applied, and pixels at the
edge of each channel were masked to improve the flatness of the background. 

A broad-band image generated from the combined MUSE data-cube is shown in Fig.~\ref{fig:img}. 
The point-spread function, measured at $\sim$7000\,\AA\ from stars in the combined observation, has a FWHM of  0.6\,arcsec (2.9\,pixels).
The tangential arc is clearly seen in the continuum image, as well as numerous other cluster and background galaxies\footnote{We have conducted a careful search for
additional lensed galaxies behind Abell 1201, which could provide improved constraints on the lensing model. 
Although many faint emission-line objects were found, none of them can be identified as multiply imaged.}.
The BCG light can be traced to $\sim$20\,arcsec ($\sim$60\,kpc).

\begin{figure*}
\includegraphics[width=168mm]{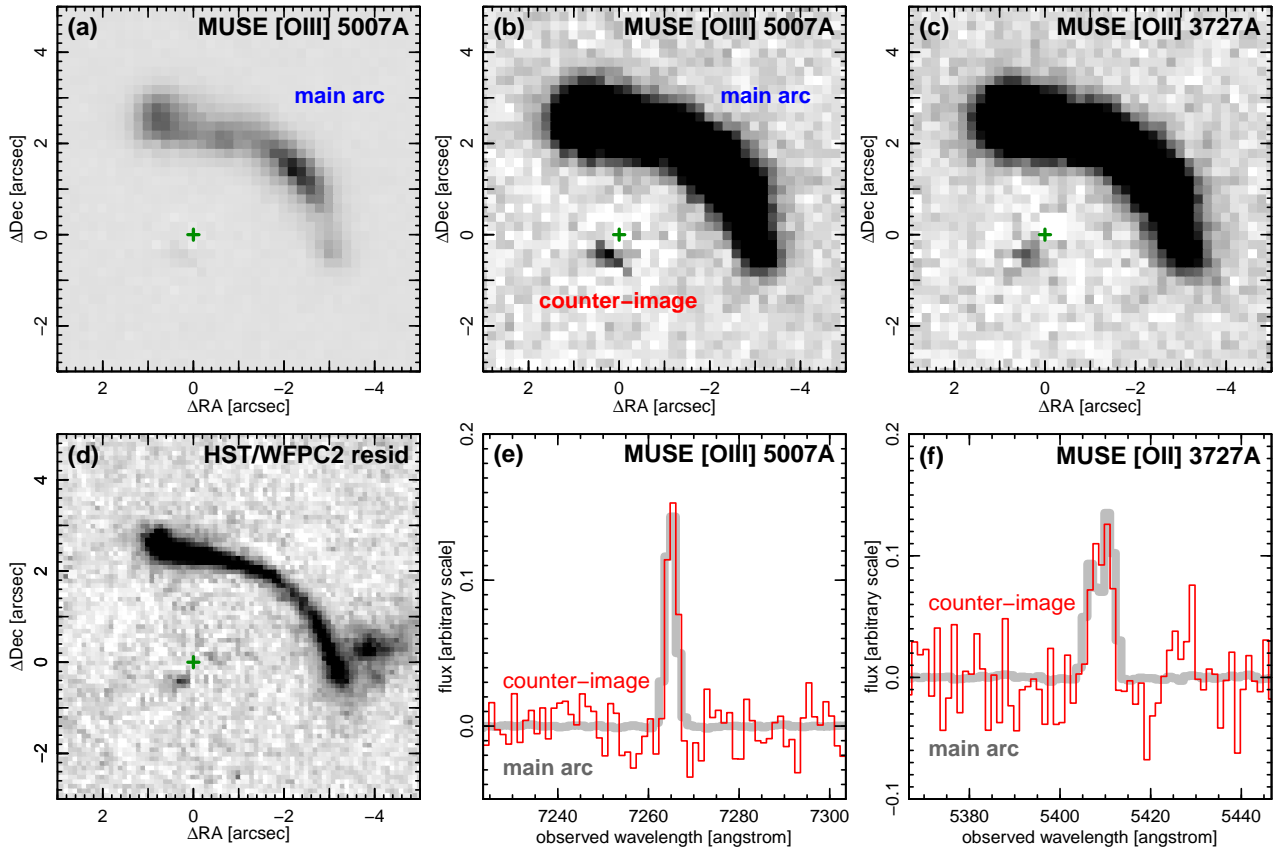}
\vskip 0mm
\caption{Discovery and confirmation of a faint counter-image to the bright arc in Abell 1201. Panel (a) shows the net  [O\,{\sc iii}] 5007\AA\ emission-line image derived from our
MUSE observations, scaled to show structure within the main arc. The green cross indicates the position of the BCG centre. 
Panels (b) and (c) show net emission-line images for [O\,{\sc iii}] 5007\AA\ and [O\,{\sc ii}] 3727\AA, with 
grey-scale emphasizing the faint peak seen close to the lens centre, which we identify as a lensed counter-image.
Panel (d) demonstrates that a peak is also visible in the {\it HST} continuum image (WFPC2/F606W), after careful subtraction of the foreground lens 
galaxy using an ellipse-fitting method. Finally, panels (e) and (f) show the extracted spectra of the main
arc and counter-image centred on the emission lines, confirming their common origin in a $z$\,=\,0.451 source.}
\label{fig:context}
\end{figure*}

Fig.~\ref{fig:context} presents the discovery of a faint counter-image to the main arc. 
Extracting the MUSE spectrum of the main arc, we find that the  [O\,{\sc iii}] 5007\,\AA\ is the brightest emission line in the MUSE spectral range. 
Integrating over this line, and subtracting a continuum derived from neighbouring wavelengths, we obtain the net emission-line image shown in Fig.~\ref{fig:context}a,b.
In addition to the main arc reported by E03, a significant excess emission is observed 
0.6\,arcsec SSE of the BCG centre, with a flux ratio $\sim$1:200, relative to the integrated value for the main arc.
A faint peak is seen at the same location in a similarly-constructed net [O\,{\sc ii}] 3727\,\AA\ image (Fig.~\ref{fig:context}c).
In fact, a residual feature is already clearly visible at this location in the E03 {\it HST} WFPC2 image, after subtracting a model for the BCG light (Fig.~\ref{fig:context}d).
Note that the bright object at $\sim$(--4,+0.5)\,arcsec is an unrelated background galaxy at $z$\,=\,0.273, which is not multiply-imaged.)
Spectra extracted at the location of the faint peak (after subtracting a model for the BCG spectrum) show that the excess flux is clearly centred on 
the expected wavelengths of the [O\,{\sc iii}] and  [O\,{\sc ii}]  lines (Fig.~\ref{fig:context}e,f). 
The weaker [O\,{\sc iii}] 4959\,\AA\ and H$\beta$ lines are not clearly detected from the inner image, but 
given the spatial coincidence of {\it HST} continuum emission with the significant emission in two lines, both of which 
are well-matched to the expected wavelengths, we consider it beyond reasonable doubt that the faint source is a lensed counter-image to the main arc. 

\begin{figure}
\includegraphics[width=85mm]{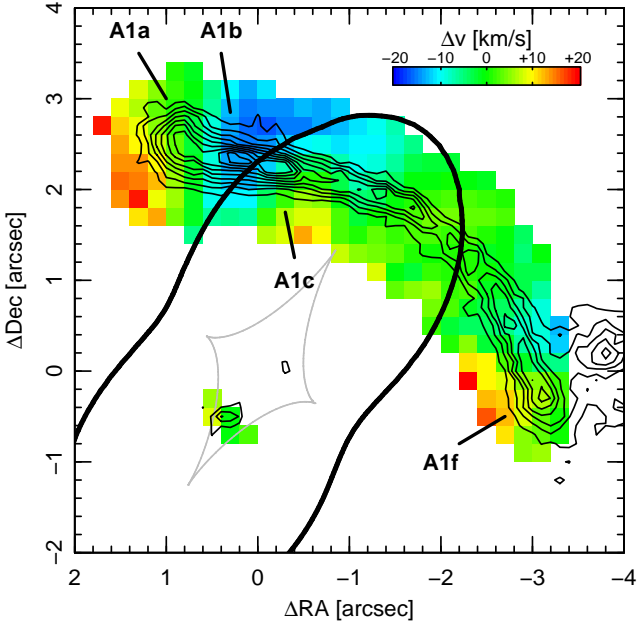}
\vskip -2mm
\caption{Velocity field in the arc, derived from a gaussian fit to the [O\,{\sc iii}] line, after applying a 3$\times$3 pixel spatial smoothing. 
Contours show the {\it HST} residual image. The thin and thick lines are the caustic and critical curves
from the lensing model in Fig.~\ref{fig:lensmods_coun}d. The labels A1a, etc, show nomenclature for local maxima introduced by E03 and referred to in the text.
}
\label{fig:arcvel_map}
\end{figure}

The spatial resolution of the MUSE data is much lower than that of the {\it HST} image, so in the lens modelling reported in 
Sections~\ref{sec:mainmod}~and~\ref{sec:count} we primarily use {\it HST}-derived positional 
constraints. However, the emission-line data can provide additional information to help verify the solutions obtained. 
Although matching the overall form of the arc as seen by {\it HST}, there are notable differences which cannot be attributed to the difference in resolution.
Strikingly, the bright image pair at $\sim$(0.0,+2.3)\,arcsec, denoted A1b/c by E03, and used by them to locate the critical curve, does not correspond to any peak in
the emission-line map. Conversely, the region of weaker continuum at $\sim$(--2.2,+1.2)\,arcsec is coincident with the {\it maximum} in the [O\,{\sc iii}] image. Other local
peaks are located at the extremities of the arc (A1a, A1f in the E03 nomenclature), roughly coincident with continuum maxima, and at
$\sim$(--0.5,+0.2)\,arcsec, which does not have a continuum counterpart. Fig.~\ref{fig:arcvel_map} shows the velocity map derived from the [O\,{\sc iii}] line.
The total velocity range is $\pm$25\,km\,s$^{-1}$. A reversal in the velocity trend along the arc, due to the lens folding, is clearly seen at $\sim$(0,+2.2)\,arcsec,
as reported previously by \citet{SwinbankThesis}.

\section{Models constrained by the main arc}\label{sec:mainmod}

E03 showed that the main arc could be reproduced by a ``cusp" configuration, where the unlensed position of the source
crosses one of the points of the tangential caustic. In this case, each point in the source maps
to three neighbouring images close to the tangential critical line; for an extended source, the images merge to form a single large
arc.

In this section, we develop models for the lens, constrained by three positions on the main arc proposed by E03 to be sister images
of one another: the A1b/c image pair bracketing the critical curve, and A1f at the far end of the arc.
The location of the velocity fold in Fig.~\ref{fig:arcvel_map} strongly supports the critical curve passing through A1b/c, {\rev though we note 
that the velocity near A1f appears discrepant with the measurements near A1b/c.}
The coordinates for the constraints, in arcsec relative to the BCG centre, are: A1b\,=\,$(-0.24,-2.34)$, A1c\,=\,$(+0.24,-2.24)$, A1f\,=\,$(+3.13,-0.26)$, determined 
by computing centroids within a 2-pixel window around each flux peak in the {\it HST} image. The estimated positional error on all constraints is 
0.025\,arcsec in each coordinate, i.e. a quarter of the {\it HST} pixel.

All lensing calculations here are made using the {\tt gravlens/lensmodel} software \citep{2001astro.ph..2340K}.

\begin{figure*}
\includegraphics[width=170mm]{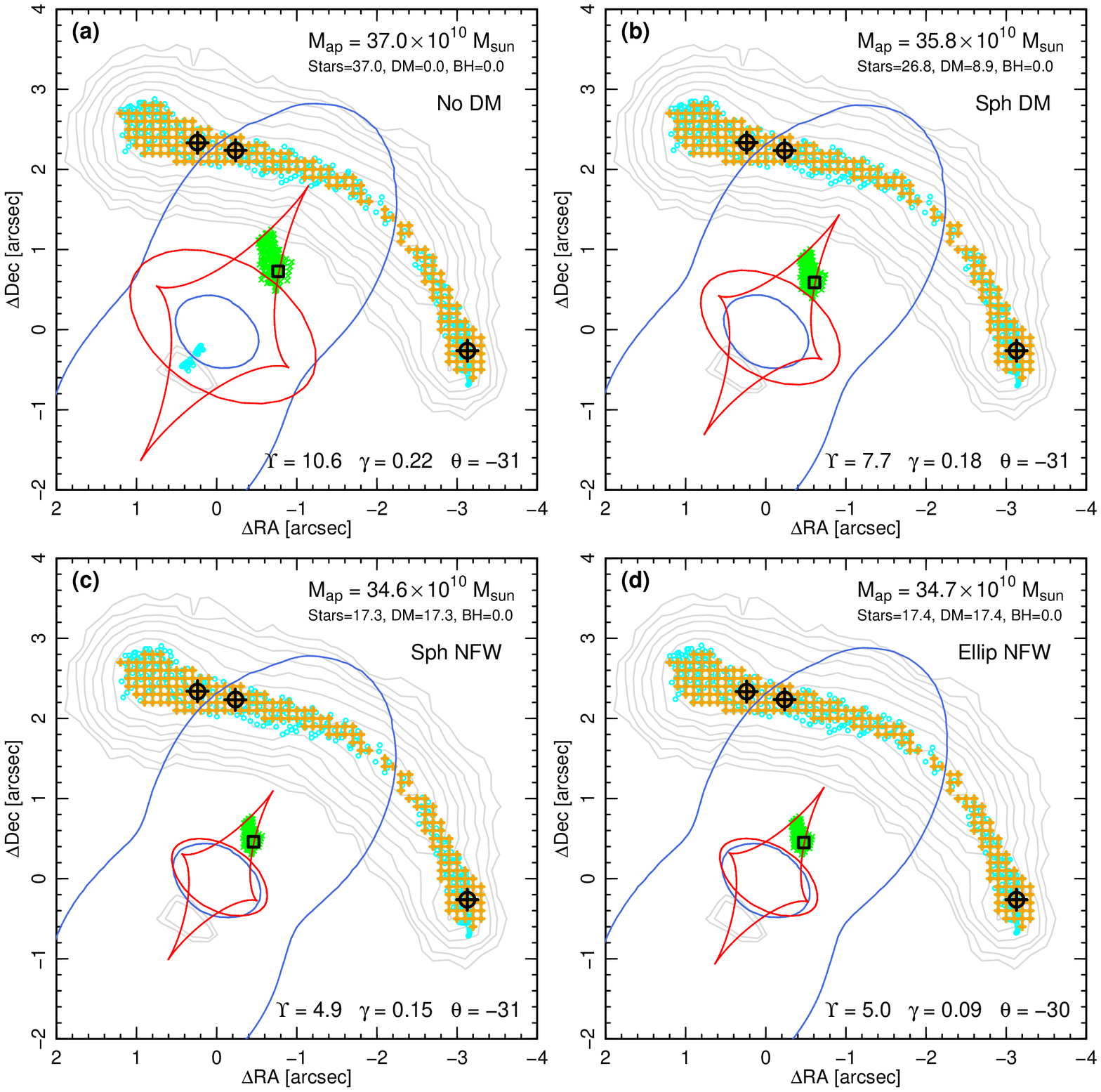}
\vskip -1mm
\caption{Lensing models for the main arc. In each panel, the blue lines show critical curves in the image plane, and the red curves are the corresponding source-plane caustics.
Heavy black circles are the input image positions used as constraints for the model optimization. Black crosses are the output predictions for these images, and the 
square marks the corresponding position in the source plane. The orange crosses show all observed pixels in the main arc, identified above a threshold in the {\it HST} image. 
These points are not used as constraints in the fitting, but provide independent validation of the models: the predicted sister images to the arc points are plotted as cyan 
circles, and their source-plane counterparts are shown in green. The grey contours show the net [O\,{\sc iii}] emission from MUSE.
For  each model we indicate the total mass projected within a circular aperture of 4.75\,kpc (1.6\,arcsec), and the contributions from stars and dark matter within this aperture.
We also note the {\it stellar} mass-to-light ratio, $\Upsilon$, and the external shear amplitude, $\gamma$, and angle, $\theta$.
Panel (a) shows the minimal model, in which all mass follows the observed light, with an external linear term. 
Panels (b) and (c) include a spherical NFW dark-matter halo component, contributing respectively 25 or 50\ per cent of the projected mass within 4.75\,kpc.
In Panel (d), the DM is assigned an ellipticity similar to that of the BCG.
}
\label{fig:lensmods_main}
\end{figure*}

\subsection{Constant $M/L$ models}

The first models we attempt to fit are those in which the mass distribution is fully determined by the observed luminosity of the BCG.

The luminosity profile of the Abell 1201 BCG exhibits a clear flattening near the centre, well beyond the radius affected by the {\it HST} PSF. 
Within the radius of the tangential arc but outside the PSF disk,  the profile is well described by an elliptical
``Nuker'' law \citep{1995AJ....110.2622L}, with outer and inner logarithmic slopes of $\beta$\,$\approx$\,1.2 and $\gamma$\,$\approx$\,0.4 respectively, and
break radius $r_b$\,$\approx$\,1\,kpc, or ``cusp radius'' $r_\gamma$\,$\approx$\,0.5\,kpc. Cores of this size are typical for BCGs 
of comparable luminosity:  at $M_V$\,$\approx$\,--23.7, \citet{2007ApJ...662..808L} derive a mean $r_\gamma$\,=\,0.45\,kpc with a factor-of-two galaxy-to-galaxy scatter.

For the lens modelling, we choose to represent the stellar mass using a pixelized convergence map, to account for the detailed profile and angular structure in the BCG.
Specifically, we construct an ellipse-fit representation of the {\it HST} image, derived after masking pixels affected by the main arc and by the $z$\,=\,0.273 source.
We calibrate to the $r$-band by matching the large-aperture flux of the BCG as measured from Data Release 8  of SDSS
\citep{2011ApJS..193...29A},
and use the source and lens redshifts to compute a lensing convergence map under an assumed  mass-to-light ratio of $M/L$\,=\,1 (in solar units). 
A scaling parameter applied to this convergence then yields the lensing estimate for $M/L$. 
For the mass-follows-light models, $M/L$ is formally identical to the {\it stellar} mass-to-light ratio $\Upsilon$, but in practice of course we expect the mass 
to include contributions from dark matter.

Without additional freedom, the constant $M/L$ model is unable to fit the three positional constraints or produce the observed arc morphology.
This is a generic result, already noted by E03: any model matched to the orientation and ellipticity of the BCG light, in which the critical curve bisects A1b/c, 
will produce a third image that does not lie on the arc. 
The simplest solution is to include greater freedom in the model, by introducing an external linear shear term, with amplitude $\gamma$ and direction $\theta$. 
The shear is intended as a first-order approximation to the effect of structure beyond the modelled region.
(E03 achieved a similar effect by adding a highly-elliptical mass component to represent the cluster potential.)
The inclusion of the shear term is motivated by the complex and asymmetric mass structure surrounding the lens:
X-ray observations and optical spectroscopy \citep{2009ApJ...692..702O,2012ApJ...752..139M} 
indicate a post-merger morphology for Abell 1201, with the merger direction aligned with the BCG major axis, 
and the BCG itself is offset from the X-ray peak by $\sim$11\,kpc
along the same axis. 

Fig.~\ref{fig:lensmods_main}a shows the lensing caustics, critical lines, and image and source positions, for the best-fitting model with constant $M/L$ and shear.
The model matches the input position constraints essentially perfectly; there is only one residual degree of freedom in the fit. 
Moreover, the model reproduces the overall morphology of the main arc: pixels in 
the arc map to one another successfully, despite {\it not} being used as constraints in the model fitting.
The arc points also map to a credible morphology in the source plane, lying across the cusp of the tangential caustic.
The emission-line structure  of the arc provides additional support for this general scenario:
the critical curve passes directly through the continuum-faint but line-bright region at $\sim$(--2.2,+1.2)\,arcsec. 
This is suggestive of an emission-line peak located near the caustic, at the northern tip of
the unlensed source, and mapping to an image pair that is unresolved at the MUSE resolution. 
The third image corresponding to this pair would be close to A1a, the peak at the north-eastern extremity of the arc, 
which is also bright in the emission map (see Fig~\ref{fig:context}a).

The derived shear is large in amplitude 
{\rev ($\gamma$\,=\,0.22)}, 
and directed 31\,deg West of North, 
in a convention where the angle indicates the direction of an external mass 
concentration generating the shear. This direction is consistent with the overall cluster axis along which the BCG and merging substructure are aligned. 
We derive a total mass-to-light of $M/L$\,=\,10.6$\pm$0.3 in $r$, where the errors are estimated by simple Monte Carlo simulation,
perturbing the positional constraints by errors of 0.025\,arcsec in each coordinate, and re-fitting the model. 
This result is consistent with the value of $M/L$\,=\,$9.4^{+2.4}_{-2.1}$ in $V$,
estimated by E03 from their two-component parametric model.
The total mass-to-light ratio is much larger than the expected value of $\sim$4 for an old stellar population with MW-like 
IMF\footnote{For example, a single-burst population formed at $z$\,$\approx$\,4, has age $\sim$10\,Gyr at $z$\,=\,0.169.
In the \cite{2005MNRAS.362..799M} models, with \cite{2001MNRAS.322..231K} IMF, this 
population has $\Upsilon$\,=\,3.5 for [Z/H]\,=\,0.0 or $\Upsilon$\,=\,4.4 for [Z/H]\,=\,+0.35. This value is for observed SDSS $r$-band, 
after correcting for a 18 per cent bandshifting effect.}, indicating
either a heavier IMF, or significant dark-matter contributions, or both. 
The projected mass inside a 1.6\,arcsec (4.75\,kpc) radius aperture is $M_{\rm ap}$\,=\,(37$\pm$1)\,$\times$10$^{10}$\,M$_\odot$.

\subsection{Models with a dark-matter halo}

We now consider models with an explicit description for the dark halo as a separate mass component, assumed to follow a 
 \citet*{1996ApJ...462..563N} (NFW) density profile.

In the cases shown in Fig.~\ref{fig:lensmods_main}, we impose a 25 or 50 per cent dark-matter fraction within 
an aperture of 4.75\,kpc radius. Panels (b) and (c) show the models with a spherical halo, while in panel (d) the halo is flattened with ellipticity $e$\,=\,0.4.
In each case, the NFW break radius is fixed at $r_s$\,=\,300\,kpc ($\sim$100\,arcsec). 
For the virial radius of {\rev 1.47\,Mpc (for our cosmology)} measured by \cite{2013ApJ...767...15R}, this corresponds to a 
halo concentration $c$\,$=$\,{\rev 4.9}, which is typical {\rev in simulations} for clusters with mass similar to that of Abell 1201\footnote{{\rev 
\citet{2007MNRAS.381.1450N} report a mean $c$\,$=$\,4.8, with a 2$\sigma$ range of 3.0--7.4, at $M_{200}$\,=\,3.5$\times$10$^{14}$\,M$_\odot$, 
for our cosmology. We have verified that even adopting $r_s$\,=\,150\,kpc, corresponding to $c$\,$\approx$\,10 makes no substantive
difference to the results in this paper.}}

After optimising the stellar mass-to-light ratio and the external shear, the models with dark matter yield fits to the main arc 
that are virtually identical to those of the mass-follows-light models. This is consistent with the expectation that only the total projected mass within the arc
would be well constrained. Adding $\sim$50 per cent dark matter naturally reduces the 
derived {\it stellar} mass-to-light ratio, to $\Upsilon$\,$\approx$\,5. 
Increasing the halo ellipticity reduces the required external shear amplitude, to 
{\rev $\gamma$\,=\,0.12}. Regardless of the form of the
adopted halo, the total projected mass within a 4.75\,kpc radius aperture is slightly reduced from the no-dark-matter case, 
with $M_{\rm ap}$\,=\,(35$\pm$1)\,$\times$10$^{10}$\,M$_\odot$.

\section{Interpretation of the counter-image}\label{sec:count}

In this section, we turn to considering models which can adequately 
predict the presence of the inner counter-image shown in Fig.~\ref{fig:context}.

\subsection{A uniformly very heavy IMF?}

In the models with dark-matter haloes (Fig.~\ref{fig:lensmods_main}b,c,d), the source lies inside a so-called ``naked cusp'', 
where the tangential caustic curve extends beyond the elliptical caustic. This unusual configuration leads to exactly three images, of similar magnification, close to 
the tangential critical line. The naked cusp arises from the combination of high ellipticity and strong shear in Abell 1201 (which determines the size of the tangential caustic),
together with the shallow total mass profile (which sets the location of the radial caustic).

In the case where all of the gravitating mass is distributed identically to the stellar light (Fig.~\ref{fig:lensmods_main}a), 
the total mass profile is steeper, so that some points in the source fall inside both caustics, and generate
additional images, forming a radial arc\footnote{\rev We use the term ``radial arc'' specifically to refer to an image generated by an extended source which
crosses the radial caustic.}.
The new images will be faint compared to the main arc (both because they are relatively demagnified by the lens, 
and because they map to outer parts of the source). Simple elliptical source models can adequately match the observed 1:200 flux ratio. 
The predicted position of the radial arc is close to, but not exactly coincident with the observed location of the counter-image, being offset $\sim$0.15\,arcsec
towards the lens centre. 

If we interpret this model as a pure ``stellar-mass'' lens, then the mass-to-light ratio of $\Upsilon$\,=\,10.6 implies an extremely heavy IMF, with a mass excess 
factor of 2.4--3.0 (relative to Kroupa), depending on the metallicity, if the stellar population is old. This is larger than the typical mass-excess factors
of 1.5--2.0 reported for giant ellipticals (see Section~\ref{sec:disc}), and of course the DM contribution is unlikely to be negligible at the centre of a massive cluster. 
We therefore consider this {\it uniformly} heavy IMF interpretation implausible on balance, and explore alternative scenarios in the following sections.

\begin{figure*}
\includegraphics[width=170mm]{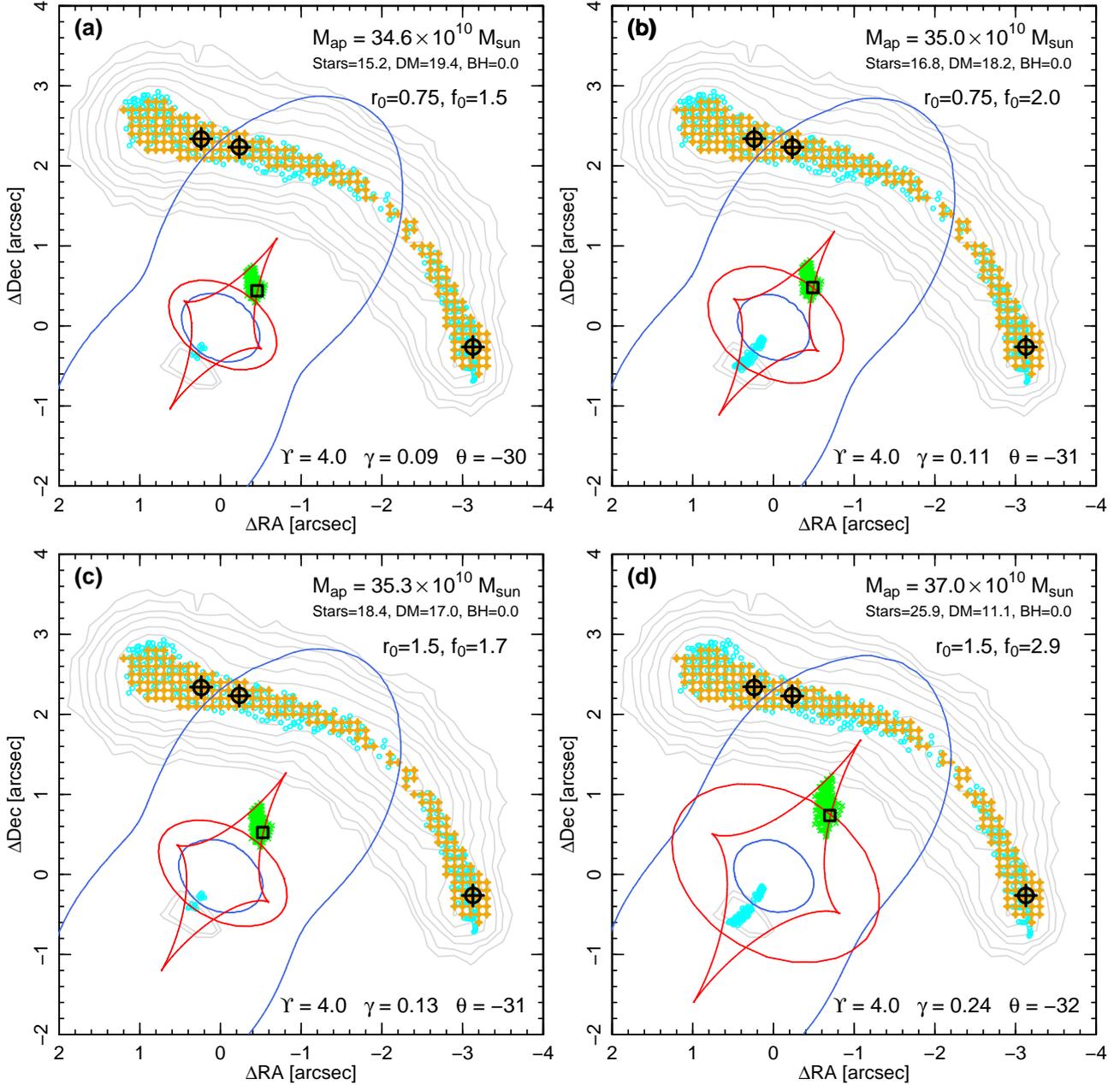}
\vskip -1mm
\caption{Examples of lensing models with radially-variable {\it stellar} mass-to-light ratio $\Upsilon$ in the central part of the BCG. A linear $\Upsilon$ 
gradient is applied within a radius of $r_0$ (0.75 or 1.5 arcsec, in the cases shown), and causes a factor of $f_0$ increase between $r_0$ and the BCG centre.
The value of $\Upsilon$ quoted in each panel is the (constant) value at radius greater than $r_0$.
 }
\label{fig:varml}
\end{figure*}

\subsection{Stellar M/L gradients?}

\begin{figure*}
\includegraphics[width=170mm]{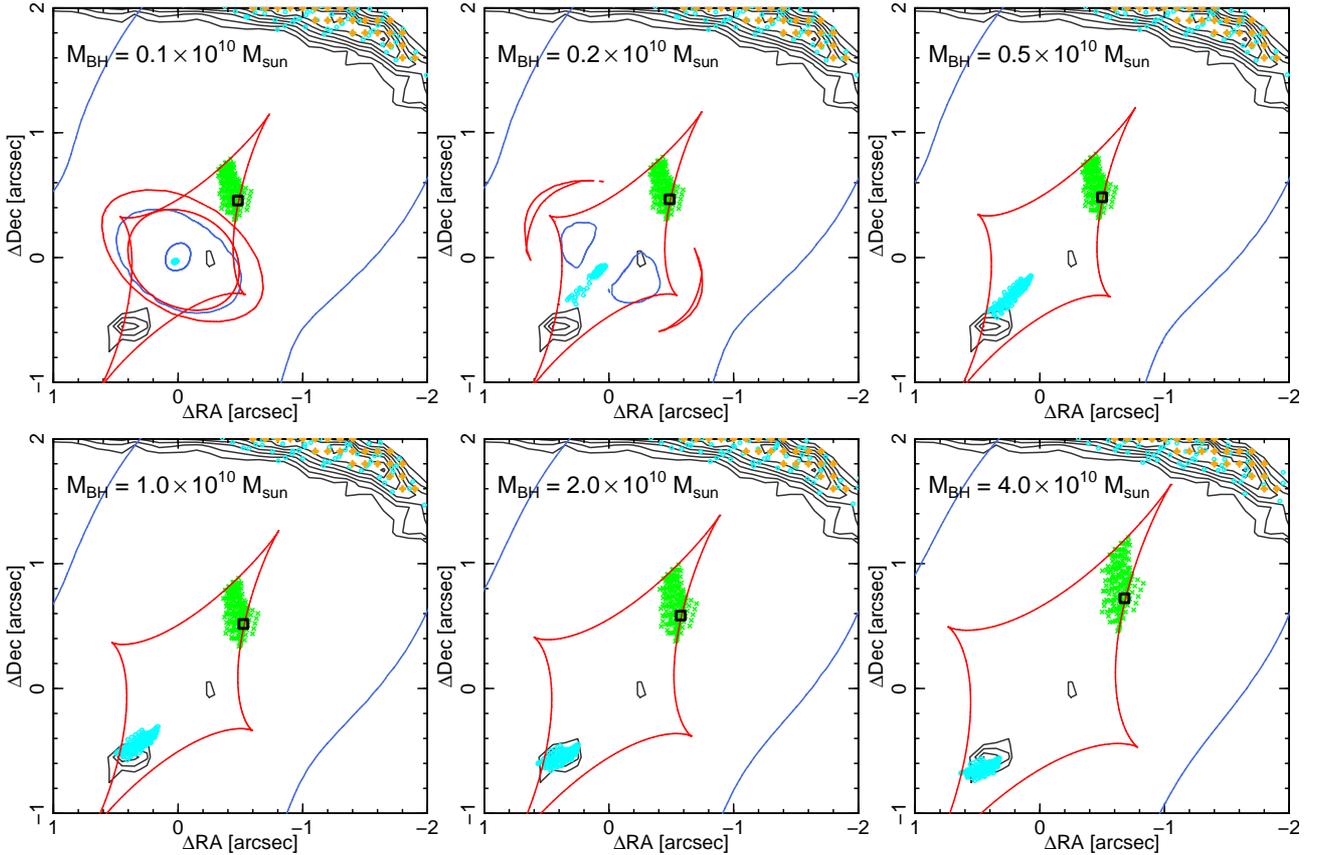}
\vskip -1mm
\caption{The effect of adding a central black hole to the model shown in Fig.~\ref{fig:lensmods_main}d. For this figure, we zoom in to the 3$\times$3\,arcsec$^2$ region
to show more clearly the behaviour of the inner counter image.
The lensing model is constrained only using the three sister images identified on the main arc, and the stellar mass to light ratio ($\Upsilon$) and
shear ($\gamma,\theta$) are re-fit in panel, assuming a fixed black hole mass which increases from panel to panel. In this figure, the grey contours 
are from the {\it HST} residual image. All of the models shown reproduce the positional constraints for the main arc, but make different predictions
for the location and flux of the inner counter-image.
Black hole masses of $\sim$2$\times$10$^{10}$\,M$_\odot$ provide the best match to the observed position,
and also yield a flux ratio comparable to the observed 1:200.
}
\label{fig:bhillus}
\end{figure*}

\begin{figure*}
\includegraphics[width=170mm]{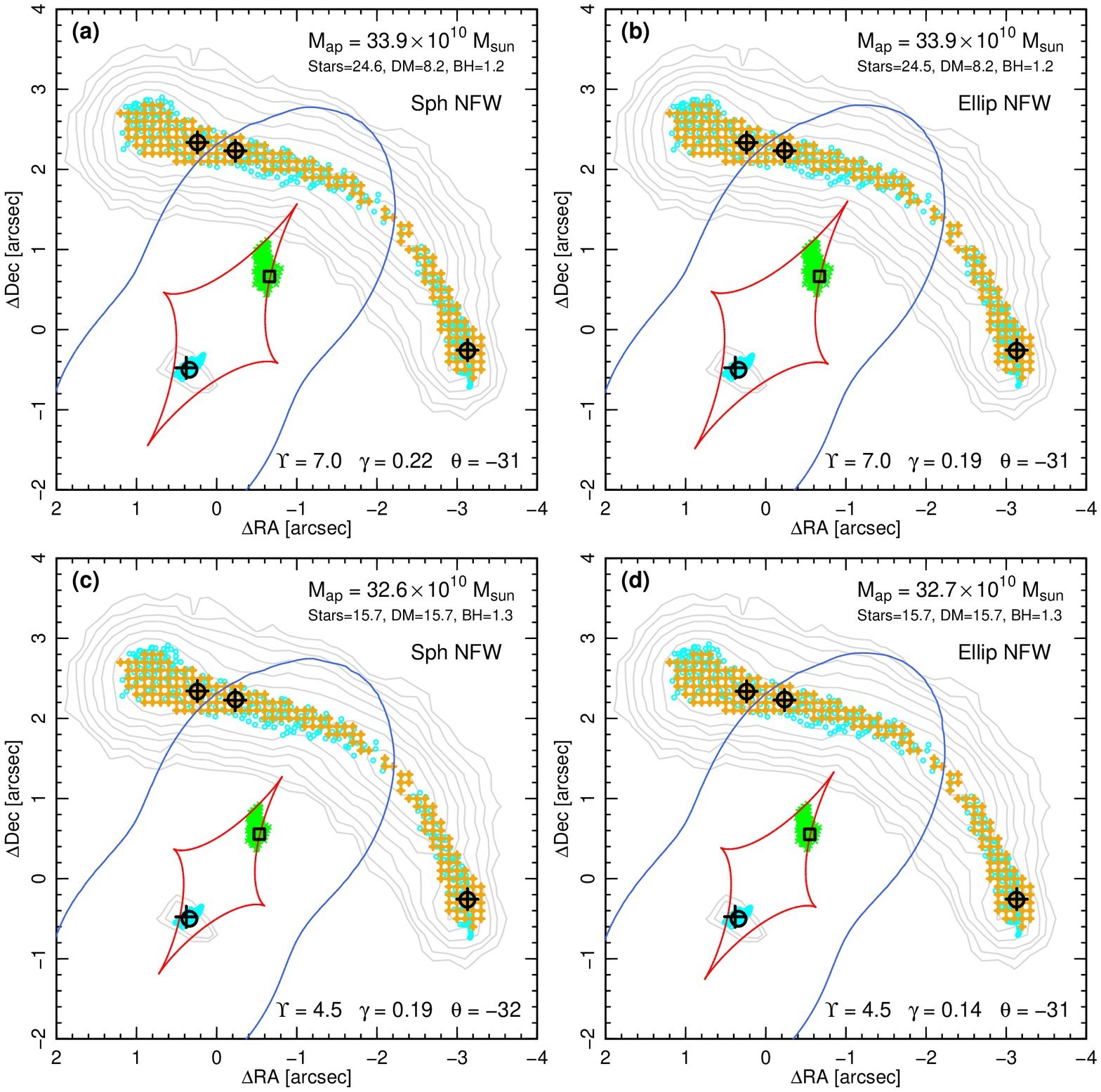}
\vskip -1mm
\caption{Lensing models incorporating a central black hole to account for the observed counter image, which is included as a fourth positional constraint.
Each panel shows the result for a different assumption regarding the dark-matter fraction and ellipticity. 
Panels (a) and (b) have stellar mass-to-light ratios corresponding to heavy (Salpeter-like) IMFs, while panels (c) and (d) correspond to MW-like IMFs.
In all cases, a black hole mass of (1.2--1.3)$\times$10$^{10}$\,M$_\odot$ is required to reproduce the counter-image.}
\label{fig:lensmods_coun}
\end{figure*}

Generically, the presence of the counter-image implies that the mass profile is at least as steep as the luminosity profile, on small scales. 
This cannot be achieved by altering the distribution of dark matter, as long as this component is flatter than the stellar profile, as expected. 
However, variations in the {\it stellar} mass-to-light ratio  $\Upsilon$, as a function of radius, could provide a way to steepen the mass profile sufficiently to 
produce the radial arc without requiring an excessively heavy IMF throughout the galaxy. 

We test this scenario by modulating the lensing convergence generated by the stars, with a linear function of slope $f_0$ 
between the galaxy centre and some threshold radius, $r_0$. The mass-to-light ratio is unchanged beyond $r_0$.
To have the desired effect, $r_0$ must be smaller than the radius probed by the main arc.
We limit our exploration to cases with $r_0$\,=1.5\,arcsec or $r_0$\,=0.75\,arcsec, and tune the dark-matter content in each model
to retain $\Upsilon$\,=\,4.0 in the outer (unaffected) region, compatible with a MW-like IMF. 

Fig.~\ref{fig:varml} illustrates several representative models with $\Upsilon$ gradients.
For each value of $r_0$, we show the case corresponding to the minimum gradient necessary to produce a
radial arc (Panels a,c), and for a larger value illustrating the effect of increasing $f_0$, with $\Upsilon$ and $r_0$ held fixed (Panels b,d). 
The main result of these tests is that
a radial arc counter-image is formed if $\Upsilon$ increases by a factor of 1.5--1.7 towards the galaxy centre, depending only slightly 
on $r_0$. 
These cases effectively add (1--4)$\times$10$^{10}$\,M$_\odot$ at small radius, relative to an assumption of constant $\Upsilon$\,=\,4.0.
At this threshold, the radial arc is similar to that in the case of the uniformly heavy IMF model, forming compact pair of images close to the critical line,
and somewhat offset from the observed counter-image location, with a flux ratio comparable to the 1:200 observed.
For steeper $\Upsilon$ gradients, a larger part of the source falls inside the quadruply-imaged region, and the radial arc  brightens  and becomes
more extended, to a length of 0.5\,arcsec or more, in contrast to the compact observed {\it HST} morphology. 

The $\ga$60 per cent variation required in $\Upsilon$ is larger than the $\sim$10 per cent attributable to typical age and metallicity gradients in massive ellipticals and BCGs
\citep{2010MNRAS.408...97K,2015MNRAS.449.3347O}. Hence this model appears to require radial gradients in the IMF; for example a factor of 1.55 in $\Upsilon$ corresponds
to the difference between a  \citet{2001MNRAS.322..231K} and an extrapolated \citet{1955ApJ...121..161S} IMF.
In summary, a model with spatial variation in the IMF directly alters the slope of the lensing potential, and consequently produces
radial-arc counter-image without requiring such {\it extreme} variation away from the MW IMF as in the spatially uniform scenario.

\subsection{A very massive central black hole?}

An alternative route to steepening the inner mass profile, without invoking a non-standard IMF at all, is to include contributions from 
a central super-massive black hole. 

The presence of a central point mass
can qualitatively alter the structure of caustics and critical lines in a lensing system, as described in detail by 
\citet{2001MNRAS.323..301M}. For the purposes of this paper, the relevant aspect is that above a critical black hole mass
(a few per cent of the total mass inside the critical curve), the usual radial caustic can be destroyed, and {\it all} source-plane positions inside the tangential caustic
become quadruply imaged. Hence the observability and location of a counter-image, for a source located in an ``otherwise-naked cusp'', can become sensitive
to the presence and mass of a central black hole. We illustrate this situation, as it applies specifically to Abell 1201, in Fig.~\ref{fig:bhillus}  
(see fig.~3 of Mao et al. for a more general description). Here, as before, the model is constrained using only the three sister-images identified on the main arc (A1b, A1c, A1f), 
and the stellar mass-to-light ratio ($\Upsilon$) and shear ($\gamma,\theta$) are re-fit in each panel, assuming a fixed black hole mass which increases from panel to panel.
For relatively small values of the black-hole mass, $M_{\rm BH}$\,$\lesssim$\,0.3$\times$10$^{10}$\,M$_\odot$, the caustic structure is complex, and 
a highly demagnified image is produced at extremely small separation ($\lesssim$\,0.1\,arcsec) from the lens centre. For more massive black holes
($M_{\rm BH}$\,$\approx$\,0.5$\times$10$^{10}$\,M$_\odot$), there is only a single caustic, and the fourth image begins to extend away from the lens centre. 
As $M_{\rm BH}$ increases further, the counter-image becomes brighter and more compact, and moves to larger separation, reaching the observed position 
for $M_{\rm BH}$\,$\approx$\,(1--2)$\times$10$^{10}$\,M$_\odot$. For the same mass, the flux ratio between main arc and counter-image is $\sim$1:100. 

{\rev In the black hole model (unlike the $\Upsilon$-gradient case), all multiply-imaged parts of the source are quadruply imaged, 
so the unresolved inner image can be assumed to be a sister image to points A1b/c/f.}
To incorporate the counter-image consistently into our modelling, we add it as a fourth positional constraint, C\,=\,$(-0.26,-0.50)$, 
and refit for four model parameters, including the black-hole mass ($\Upsilon,\gamma,\theta,M_{\rm BH}$).
As before, the fit does not distinguish between dark matter and stellar mass, so we test 
the results using different fixed assumptions for the halo shape and mass contribution. 
Four illustrative cases are shown in Fig.~\ref{fig:lensmods_coun}.
For each form assumed for the halo, we find that the presence and position of the counter-image are reproduced for black-hole masses of (1.2--1.3)$\times$10$^{10}$\,M$_\odot$.
Adopting the flattened-halo case with $\sim$50 per cent halo contribution (Fig.~\ref{fig:lensmods_coun}d)   as 
our default solution, and using Monte Carlo simulations as before to propagate positional errors, we 
obtain a mass of  $M_{\rm BH}$\,=\,$(1.3^{+0.6}_{-0.5})$\,$\times$10$^{10}$\,M$_\odot$. 
In this model, the stellar mass-to-light ratio is $\Upsilon$\,=\,4.5$\pm$0.3, consistent with a Milky-Way-like IMF.
Reducing the dark-matter fraction leads to slightly more stellar mass in the galaxy centre, and hence less mass is allocated to the 
black hole. This is a small effect, however: halving the dark-matter content reduces the derived $M_{\rm BH}$ by only $\sim$10 per cent
(Fig.~\ref{fig:lensmods_coun}b). Finally, we note that in these models the total mass projected within 4.75\,kpc is 
$M_{\rm ap}$\,=\,(33$\pm$2)\,$\times$10$^{10}$\,M$_\odot$, including contribution from the black hole itself ($\sim$4 per cent of the total).
The aperture mass is consistent with that derived for the models with stars and DM only.

\section{Discussion}\label{sec:disc}

We have proposed three possible interpretations for the newly-discovered counter-image to the Abell 1201 arc: (a) a very heavy IMF throughout the BCG, 
(b) a steep increase in the stellar mass-to-light ratio towards the galaxy centre, or (c) a very massive central black hole. In this section, we assess the plausibility of 
each scenario with reference to evidence from other studies, and discuss possible routes to distinguishing observationally between the possibilities.

Scenarios (a) and (b) both require variations in the stellar IMF away from the form pertaining apparently almost universally within the Milky Way 
\citep*{2010ARA&A..48..339B}. The evidence for heavier\footnote{Either bottom-heavy with an excess of dwarf stars, or top-heavy with an
excess of remnants, relative to the Milky Way case. Both lead to larger stellar mass-to-light ratios.}  IMFs in massive galaxies has been 
discussed widely in recent years. 
From dynamical modelling of nearby early-type galaxies, \citet{2013MNRAS.432.1862C} found a
trend of increasing IMF mass factor, from MW-like ($\alpha$\,$\approx$\,1) at $\sigma$\,=\,100\,km\,s$^{-1}$ to Salpeter-like ($\alpha$\,$\approx$\,1.6) at 
$\sigma$\,=\,300\,km\,s$^{-1}$. 
\citet{2010ApJ...709.1195T} combined stellar dynamics with strong lensing for the SLACS (Sloan Lensing Advanced Camera for Surveys) lens sample, 
and derived larger mass excesses, $\alpha$\,$\approx$\,2, for the most massive galaxies ($\sigma$\,\ga\,300\,km\,s$^{-1}$), under the assumption
of universal NFW haloes. Conversely, for a sample of three very nearby strong-lensing ellipticals with $\sigma$\,\ga\,300\,km\,s$^{-1}$ 
(subject to smaller corrections for dark matter), \citet{2015MNRAS.449.3441S} found $\alpha$\,$\approx$\,1, from a pure lensing analysis, i.e.
with no dynamical inputs. Independently, the strength of gravity-sensitive features in the spectra of massive galaxies suggests they harbour
an excess of dwarf stars compared to a MW-like IMF, leading to higher mass-to-light ratios, e.g. $\alpha$\,=\,1.5--2.0 from \citet{2012ApJ...760...71C}. 
However, direct comparisons of these results to $M/L$ measurements are hampered by the unknown 
detailed {\it shape} of the IMF at very low mass \citep{2016MNRAS.463.3220L}.

The mass excess factor of $\alpha$\,=\,2.4--3.0 (relative to the Kroupa IMF), required by our spatially-uniform IMF model for the Abell 1201 counter-image,
is substantially larger than the factors discussed in the recent literature. Moreover, this scenario would imply that dark matter contributes negligibly within
the aperture probed by the main arc, contrary to theoretical expectation. For example, extracting average profiles of massive galaxies 
($M_*$\,$>$\,$10^{11}$\,M$_\odot$, $\sigma$\,$>$\,250\,km\,s$^{-1}$) from 
the 
EAGLE\footnote{Evolution and Assembly of GaLaxies and their Environments.}
simulation \citep{2015MNRAS.446..521S}, we find a typical  dark-matter mass of 
$\sim$10$\times$10$^{10}$\,M$_\odot$ projected within 4.75\,kpc, i.e. about a third of the total lensing mass. 
In rich clusters (which are not well represented in the simulation dataset) the DM fraction may well be larger.
{\rev Alternatively, integrating an NFW halo with $c$\,=\,5, $M_{200}$=\,3.9$\times$10$^{14}$\,M$_\odot$ and $R_{200}$=\,1.5\,Mpc
\citep{2007MNRAS.381.1450N,2013ApJ...767...15R}, yields 
$\sim$20$\times$10$^{10}$\,M$_\odot$}
projected within
4.75\,kpc (two thirds of the lensing mass).
On balance, we consider scenario (a) to be the least plausible of the three interpretations. 

Scenario (b), invokes an internal {\it gradient} in the stellar mass-to-light ratio $\Upsilon$, spanning a factor of $\ga$1.6 between the BCG centre and a radius of $\sim$1.5\,arcsec.
The typical change in [Z/H] over this radial interval in nearby ellipticals is $\la$0.2\,dex \citep{2010MNRAS.408...97K}, which would yield a $\la$10 per cent
effect in $\Upsilon$, according to the \citet{2005MNRAS.362..799M} models, while the age profiles are generally flat. 
\citet{2015MNRAS.449.3347O} report even shallower metallicity gradients for a sample of BCGs\footnote{The Abell 1201 BCG itself does not seem to have
unusual radial trends compared to these larger samples. Fitting SSP model spectra from \citet{2012ApJ...747...69C} to spectra extracted from 
annuli in the MUSE datacube, assuming common age in all bins, we find very weak metallicity gradients in the inner 2\,arcsec. Allowing age to vary, we tentatively 
find an increase of $\Upsilon$ towards the galaxy centre, but only by $\sim$25 per cent.}. Hence, radial variation in the IMF is probably
required to generate the necessary $\Upsilon$ gradient. 

The IMF gradient scenario requires only a fairly modest deviation from the standard MW form (at least as compared to
what is needed for the uniformly-heavy IMF proposal). Even in the galaxy centre, the mass excess factor can be as small as $\alpha$\,$\approx$\,1.6, which is 
similar to a Salpeter IMF, and more consistent with the results obtained from other methods, e.g. gravity-sensitive spectral features.
The possibility of IMF gradients has been advanced as 
a potential explanation for discrepancies between spectroscopic analyses (often limited to the innermost parts of galaxies, $R$\,$\la$0.1\,$R_{\rm eff}$) and 
dynamical measurements ($R$\,$\la$1\,$R_{\rm eff}$)  \citep{2014MNRAS.443L..69S}.
Several recent studies have attempted to measure internal gradients in the IMF from spectral features. 
For two massive galaxies, \citet{2015MNRAS.447.1033M} and \citet{2016MNRAS.457.1468L}  report steep radial trends in the derived IMF slope within 
0.5\,$R_{\rm eff}$; for their preferred description of the behaviour at very low stellar masses, the variation corresponds to a factor of 1.5--2.0 in $\alpha$.
{\rev Similar results have been reported very recently for six galaxies by \citet{2016arXiv161109859V}.}
Other studies, however, have argued that while steep gradients in some spectral features are indeed present, the pattern of trends (especially the weakness or absence of 
gradient in the Fe\,H Wing--Ford band) is more consistent with abundance variation than with IMF trends \citep*{2016ApJ...821...39M,AltonSubmitted}.
For the Abell 1201 BCG itself, the classic IMF indicators (e.g. the Na\,{\sc i} 8200\,\AA\ doublet) are redshifted out of the MUSE spectral range, 
so we cannot make a direct comparison between methods at this stage. 
Meanwhile, \citet{2016MNRAS.tmp.1467D} have analysed the dynamics of molecular gas disks in seven 
early-type galaxies, finding no clear cases of a central rise in $\alpha$ (though selection by presence of molecular gas inevitably
biases the sample away from old, very massive ellipticals). 
Hence, while there is not yet a secure consensus regarding gradients in the IMF, the magnitude of the trend required by our lensing models
of the Abell 1201 BCG is within the range being discussed in the literature, and we conclude that scenario (b) is not inconsistent with external evidence.

\begin{figure}
\includegraphics[width=85mm]{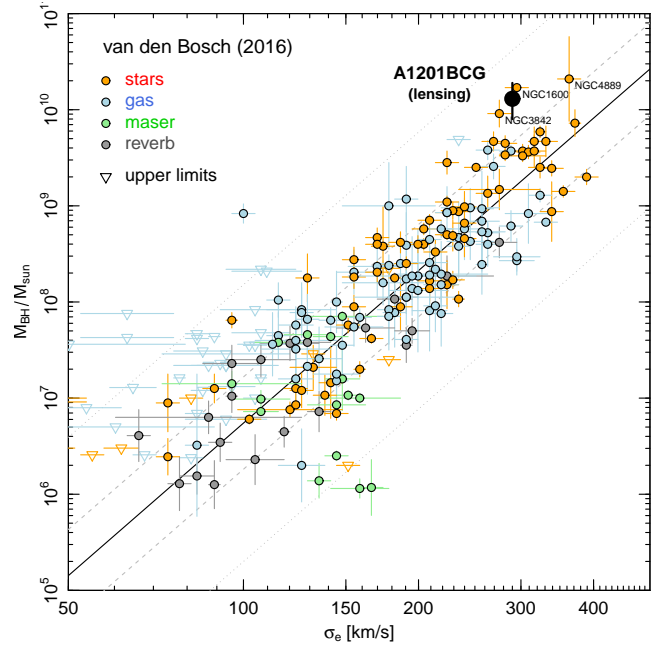}
\caption{The lensing result for Abell 1201 in comparison to the black-hole mass versus velocity dispersion relation, from the compilation of 
van den Bosch (2016).
}
\label{fig:remco}
\end{figure}

\begin{figure*}
\includegraphics[width=170mm]{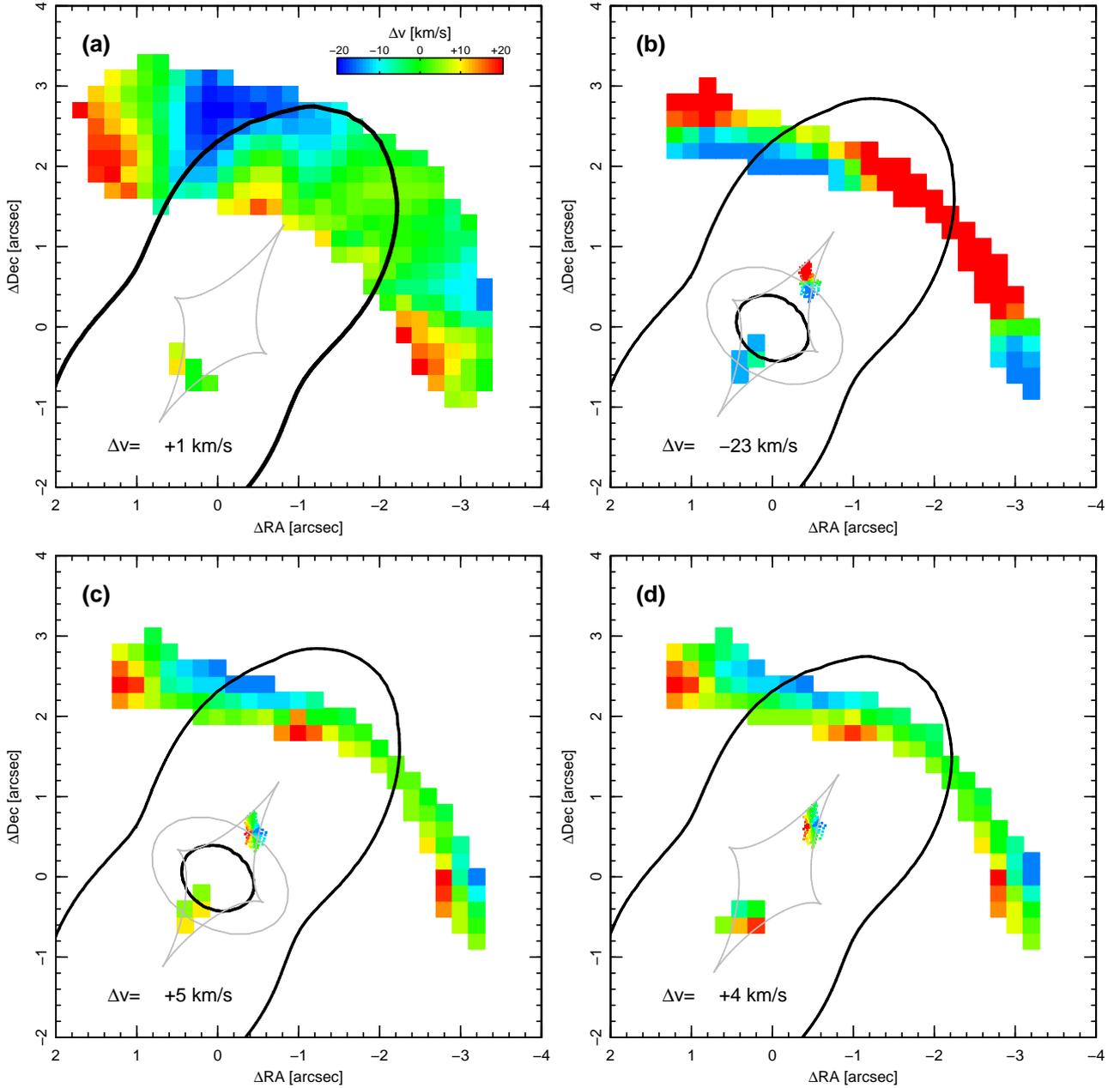}
\vskip -1mm
\caption{Observed and predicted image-plane velocity maps. 
Panel (a) reproduces the {\it observed} velocity map from Fig.~\ref{fig:arcvel_map}
To create the prediction in Panel (b), we adopt the IMF-gradient lensing model from Fig.~\ref{fig:varml}b, and impose a velocity gradient in the 
source plane, constructed to mimic a disk-like rotation in the source. The coloured regions show how this velocity field maps through to the
MUSE pixels in the image plane. The set-up in Panel (b) produces a very poor match to the MUSE observations for the main arc. 
In particular, the predicted western critical-curve-crossing corresponds to an extreme of the velocity field (red), whereas
the observed velocities are close to the mean (green) at this point. In Panel (c), we rotate the intrinsic velocity field by 90 degrees, keeping the same 
lensing model. The image-plane velocities now show a plausible semblance to the observed configuration, modulo resolution differences, 
{\rev but we note that the source-plane velocity structure is not consistent with a simple rotating disk in this case.}
Panel (d) shows an equivalent case for the black-hole lensing model from Fig.~\ref{fig:lensmods_coun}c, with the same source-plane velocity field as Panel (c).
The mean velocity offset of the counter-image from the main arc is not a good discriminant between lensing models if this velocity field is assumed.
}
\label{fig:velpred}
\end{figure*}

Finally, scenario (c) can account for the counter-image without {\it any} modifications to the IMF, by introducing a very 
massive central super-massive black hole.
Empirically, black hole masses correlate most strongly with the velocity dispersion \citep{2012MNRAS.419.2497B}, with the largest reported 
masses from dynamical studies being (1--2)$\times$10$^{10}$\,M$_\odot$, e.g. in 
the BCGs NGC\,3842 and NGC\,4889 \citep{2011Natur.480..215M} and in the field elliptical NGC\,1600 \citep{2016Natur.532..340T}. 
To place the putative Abell 1201 black hole on the scaling relations established from more traditional methods, 
we use {\tt ppxf}  \citep{2004PASP..116..138C} to  measure $\sigma_{\rm e}$, the velocity dispersion within the effective radius of $R_{\rm eff}$\,=\,5\,arcsec, 
from the MUSE datacube, finding $\sigma_{\rm e}$\,=\,290$\pm$1\,km\,s$^{-1}$.
This  is substantially larger than the values reported by \cite{2004ApJ...604...88S}, who measured
$\sigma$\,=\,230--250\,km\,s$^{-1}$ within 1.5\,arcsec. 
The discrepancy is due partly to a real mismatch\footnote{
\citet{2013ApJ...765...24N} reported similar disagreements with the Sand et al. measurements for 
several other clusters, and concluded that the earlier data were compromised by poor stellar templates and measurement procedures 
(see their section 6.4).} 
in the inner regions (where we measure $\sigma$\,$\approx$\,270\,km\,s$^{-1}$ from MUSE, and SDSS reports 277$\pm$14\,km\,s$^{-1}$),
and partly to a rising velocity dispersion profile beyond $\sim$2\,arcsec.

Fig.~\ref{fig:remco} shows our estimate for Abell 1201 BCG in comparison to the $M_{\rm BH}$--$\sigma_{\rm e}$ relation derived by \cite{2016ApJ...831..134V}
from a compilation of published masses. The predicted mean black-hole mass at $\sigma_{\rm e}$\,=\,290\,km\,s$^{-1}$ is 
$\sim$1.5$\times$10$^{9}$\,M$_\odot$, an order of magnitude smaller than the lensing-derived value. Given the large scatter around the 
relationship, the Abell 1201 BCG is a 2$\sigma$ outlier, comparable to NGC\,3842 or NGC\,1600.
The possibility of a BH mass offset in BCGs has been discussed in the context of AGN feedback in cluster environments.
\citet{2012MNRAS.424..224H} argue that the radio/X-ray properties of BCG nuclei in cool-core clusters 
are inconsistent with the ``fundamental plane'' of nuclear accretion activity \citep*{2003MNRAS.345.1057M}, 
if their BH masses follow the local scaling relations. The discrepancy can be resolved if $M_{\rm BH}$ is
under-predicted by the correlations with $\sigma$ or luminosity, by an order of magnitude.  
The direct evidence for such over-massive BHs in BCGs remains sparse, given the small number of such galaxies having dynamical mass estimates;
our measurement for Abell 1201, if confirmed, would add anecdotal support for this picture.

Which model is favoured, given these external considerations?
At a qualitative level, the existence of central super-massive black holes is established beyond reasonable doubt, and the
presence of 10$^{10}$\,M$_\odot$ objects in some of the most massive galaxies is supported by other studies.
The reality of IMF variations in such galaxies is not yet, in our view, confirmed to the same degree of confidence. 
However the {\it level} of variation required, if applied as a gradient within the BCG, is fairly modest, with a Salpeter-like
IMF in the centre sufficient to match the observed configuration. Weighing the arguments above, we mildly favour scenario (c) (a massive black hole)
over scenario (b) (an internal IMF gradient), and strongly disfavour only scenario (a) (a heavy IMF throughout). 

\section{Observational outlook}\label{sec:outlook}

We now discuss possible observational tests to discriminate better between the two remaining explanations for the counter-image.

 {\rev  From a purely lensing perspective, deeper and better-sampled {\it HST} imaging will help by 
revealing the morphology of the counter-image, which cannot be adequately established from the present very shallow WFPC2 data. 
If the counter-image is a true radial arc (i.e. formed by a an extended source crossing the radial caustic), then we expect in general to 
observe a more radially-extended image, or an image pair around the corresponding critical line. As shown in Figure~\ref{fig:varml},
the length of the arc in this case carries information about the slope of the IMF gradient.  Time has been allocated for future
observations with WFC3 (Wide Field Camera 3) in two bandpasses, with one bluewards of the BCG 4000\,\AA\ break, to maximise contrast of the arc
against the lens.
These two-colour {\it HST} observations will also provide a link which helps understand the differences between
the continuum flux (at high spatial resolution) and the emission-line structure of the main arc (only available at ground-based resolution).}

A possible route to excluding the black hole scenario is that in this model the {\it whole} of the source 
is quadruply imaged, while in the other options the counter-image is formed from only a small part of the background galaxy.
Hence by measuring properties with spatial structure within the background galaxy, it might be possible to ``tag'' the counter-image to 
a particular region in the source, if scenario (b) is correct. Radial velocity would be the most obvious tagging parameter, and naively 
the elliptical distribution of the main-arc pixels, when mapped to the source plane, suggests that the radial arc should correspond to an extreme of 
the rotation curve. If so, a velocity offset of the counter-image, relative to the mean velocity for the main arc, would decisively favour the radial arc interpretation.
In practice, however, the observed velocity structure in the arc seems to indicate a velocity gradient {\it orthogonal} to the expected 
direction\footnote{{\rev Using pixelised source reconstruction methods on the forthcoming WFC3 data will help to establish the origin of this velocity gradient, 
e.g. a merger or other peculiarity.}} (Fig.~\ref{fig:velpred}), so that no large velocity offset is predicted for the counter-image.
At higher spatial resolution, e.g. IFU with adaptive optics (AO), the internal velocity structure of the counter-image might be able to 
distinguish a reduced image of the complete source versus a small section, but the small velocity width and faintness of the image will make this
very challenging. Alternatively other possible spatially-varying ``tags'', such as emission line 
ratios or equivalent widths, could be explored, either with AO-assisted IFU data or with {\it HST} narrow-band imaging.

Independent of the information available from lensing, analysing the spatially-resolved stellar kinematics will help to establish the 
mass contribution of dark matter, through its increasing importance at large radii. Moreover, despite the large distance of the
Abell 1201 BCG compared to the other galaxies in Fig.~\ref{fig:remco}, a 10$^{10}$\,M$_\odot$ black hole would have measurable effects on the kinematics
at small radius, even in natural-seeing observations. An analysis of the kinematic data from our current MUSE observations will be presented in 
a forthcoming paper. Future AO-assisted IFU observations could provide an unambiguous dynamical confirmation of the black hole, if 
it is really as massive as required by our scenario (c).

\section{Conclusions}\label{sec:concs}

We have presented new deep IFU observations of the lensing BCG of Abell 1201, revealing a faint inner counter-image to the previously-known bright arc. 
Our lens modelling shows that if the lensing mass is dominated by a reasonable combination of stars (with constant mass-to-light ratio) 
and halo dark matter, then the bright arc arises from a ``naked cusp'', and no counter-image is predicted.
To accound for the newly-discovered image, we find that additional mass is required at the centre of the BCG, relative to the observed luminosity profile.

If the central mass distribution is dominated by stars and dark matter, then the counter image must be interpreted as a radial arc. This would require either: 
\begin{itemize}
\item
A negligible dark-matter contribution at the centre of this massive cluster, and hence a high stellar mass-to-light ratio. This would require a very heavy IMF,
with a mass-excess factor of $\alpha$\,$\ga$\,2.4 (relative to a MW-like IMF), which is larger than the typical factors of $\sim$1.6 obtained from lensing, dynamical
and spectroscopic studies of giant ellipticals \citep{2010ApJ...709.1195T,2012ApJ...760...71C,2013MNRAS.432.1862C}. 
Or: 
\item
A steep gradient in the stellar mass-to-light ratio within the radius of the main arc, increasing towards the BCG centre. Such a gradient could be generated by a 
radial trend in the IMF; in this case only modest variations away from the MW form are necessary (reach $\alpha$\,$\approx$\,1.6 at the centre), in line with some 
estimates from spectroscopy \citep{2016MNRAS.457.1468L, 2015MNRAS.447.1033M}.
\end{itemize}
Alternatively, if the stellar populations conform to a standard MW-like IMF throughout the BCG, then the observed counter-image can be reproduced by a third scenario:
\begin{itemize}
\item
A 1.3$\times$10$^{10}$\,M$_\odot$ central black hole, comparable to the most massive black holes measured from stellar dynamical modelling, 
and an order of magnitude larger than the prediction from the $M_{\rm BH}$--$\sigma_{\rm e}$ scaling relation. 
\end{itemize}

The black hole  model and the IMF variation models make differing predictions for the morphology of the counter-image.
Hence, improved high-contrast and high-resolution observations may be able to distinguish between the competing interpretations.
In a forthcoming paper, we will present measurement for the extended stellar kinematics of the Abell 1201 BCG, to derive independent constraints on the 
stellar, dark-matter and black-hole mass components, from dynamical modelling.

\section*{Acknowledgements} 
RJS acknowledges support from the STFC through grant ST/L00075X/1.
We thank Mark Swinbank for helpful conversations about MUSE data reduction and previous observations of Abell 1201, 
and Tom Collett for discussions about the lensing analysis. The datasets used in this paper are available from the {\it HST} and ESO data archives.

\bibliographystyle{mnras}
\bibliography{rjs} 

\bsp	
\label{lastpage}
\end{document}